\documentclass[twocolumn,showpacs,preprintnumbers,amsmath,amssymb]{revtex4}

\usepackage{graphicx}
\usepackage{dcolumn}
\usepackage{bm}

\begin{document}

\title{The Semi-Quantum Computer}

\author{Reinaldo O. Vianna}
\email{reinaldo@fisica.ufmg.br}
\author{Wilson R. M. Rabelo}
\author{C. H. Monken}
\affiliation{Universidade Federal de Minas Gerais - Departamento  de F\'{\i}sica\\
Caixa Postal 702 - Belo Horizonte - MG -  Brazil - 30.123-970}
\date{\today}

\begin{abstract}
  
 We discuss the performance of the Search and Fourier Transform algorithms on a
hybrid computer constituted of classical and quantum processors working together.
We show that this {\em semi}-quantum computer would be an improvement over a pure
classical architecture, no matter how few qubits are available and, therefore, it
suggests an easier implementable technology than a pure quantum computer with 
arbitrary number of qubits.

\end{abstract}

\maketitle

\section{Introduction}

In the 1980's, Feynman\cite{Feynman} suggested that computers based on the laws of Quantum 
Mechanics could be more efficient than the classical ones. In the 1990's, the discovery of
Shor's\cite{Shor} and Grover's\cite{Grover} algorithms stimulated the search for the physical 
implementation of a quantum computer. Diverse architectures have been tried\cite{Review} and, among the most sucssesfull, we can cite Nuclear
Magnetic Ressonance (NMR), Ion-traps, Optics\cite{Optics}. 
So far, the architecture that has achieved the largest number of qubits, namely seven, is NMR\cite{NMR-7}. The difficulty of implementing arbitrary number of qubits has motivated alternative 
approachs to quantum computing. It has been demonstrated that classical optics supports quantum
computing, but the number of optical elements needed grows exponentially with the number of 
qubits\cite{Classical-optics}.
Semi-classical architectures of quantum computing have also been suggested, with experimental 
implementations of the Fourier Transform of Shor's  algorithm\cite{semi-classico-QFT}, and the
Grover's search algorithm\cite{semi-classico-Grover}. The kernel of these proposals is to simulate
the two-qubit gates by use of classical communication among quantum gates. A quite different
approach, which is a kind of quantum/classical hybrid distributed computation,  dubbed type II quantum computation, has been adopted by Yepez\cite{Yepez}. A different kind of distributed quantum computation has also been suggested by other authors\cite{distributed-QC}.

\begin{figure}
\includegraphics{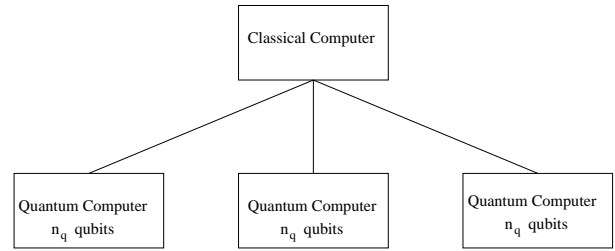}
\caption{\label{fig1} The semi-quantum computer.}
\end{figure}

The semi-quantum computer, or type II quantum computer according to Yepez\cite{Yepez}, 
is in part quantum and in part classical. The quantum part is a set of individual
genuine quantum computers with $n_q$ qubits each. These quantum nodes communicate classically 
with a classical computer, which may include several classical processors in 
parallel, as well [Fig.\ref{fig1}].   The key point here is that the quantum nodes are
not intended to solve a whole algorithmic problem, but just a part, a subroutine,
consistent with their supported number of qubits and coherence time. This hybrid
architecture explores both quantum and classical parallelisms simultaneously.
The quantum nodes send their partial results to the classical computer which, by 
its turn, performs some data processing and feeds back the quantum nodes with new
tasks, which could be the same operation as before but on a new quantum state.

Our intention here is to look for evidences that tell us if this hybrid architecture
could result in a computer that is faster than the classical ones. 
In order to do so, we are going to discuss how the Search\cite{Grover} and 
Fourier Transform\cite{Shor} quantum algorithms could be adapted to the
semi-quantum computer and investigate, through a complexity analysis, how
they perform as compared to their pure classical and pure quantum counterparts.
In section II, we discuss the Search problem. Section III is dedicated to the
Fourier Transform. We conclude in section IV.

\section{The Search Problem}

Supose that a list of $N$ elements has to be searched, in order to find the 
$M$ solutions ($M<N$) of a given problem, and that each solution can be 
identified by an appropriate oracle. Classically, all the $N$ 
elements have to be tested and, therefore, this procedure costs $O(N)$ operations. 
Grover showed\cite{Grover} that searching quantum mechanically costs just 
$O(\sqrt{N})$, a quadratic improvement over the classical algorithm.

The quantum algorithm is as follows. Associating an orthonormal basis 
$\{ |x\rangle, \,\,x=0,1,...,N-1\}$ to the indices of the elements in
the list, define a state vector formed by the equal superposition of
these basis states,

\begin{equation}
\label{Eq1}
|\psi\rangle = \frac{1}{\sqrt{N}}\sum_{x=0}^{N-1}|x\rangle\,\,.
\end{equation}

\begin{figure}
\includegraphics{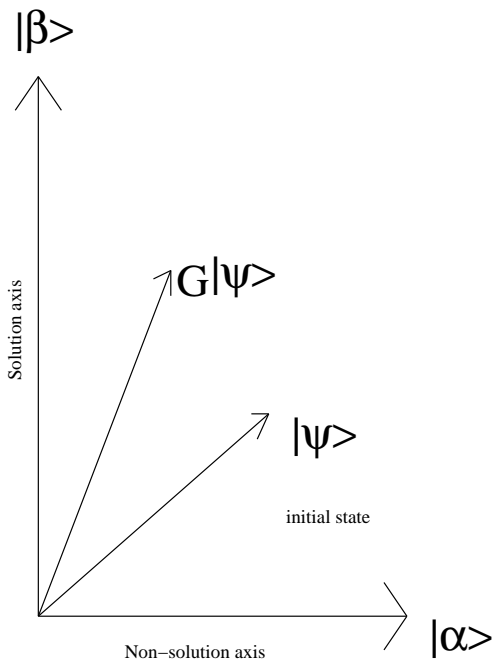}
\caption{\label{fig2} The Grover algorithm.}
\end{figure}

Consider a bi-dimensional space state spanned by the states $|\alpha\rangle$ and
$|\beta\rangle$, which are the superposition of all the non-solutions,
\begin{equation}
\label{Eq2}
 |\alpha\rangle = \frac{1}{\sqrt{N-M}}\sum_{x=non-solution}|x\rangle 
\end{equation}
and the superposition of all the solutions,
\begin{equation}
\label{Eq3}
 |\beta\rangle = \frac{1}{\sqrt{M}}\sum_{x=solution}|x\rangle\,\,.
\end{equation}
Therefore $|\psi\rangle$ can be rewritten as
\begin{equation}
\label{Eq4}
 |\psi\rangle = |\alpha\rangle\langle\alpha|\psi\rangle + |\beta\rangle\langle\beta|\psi\rangle\,\,.
\end{equation}
Now build an operator $G$, based upon the oracle for the problem (see \cite{Grover} for details), 
whose action over $|\psi\rangle$ is to increase its projection on $|\beta\rangle$ [Fig.\ref{fig2}].
After applying $G$ a sufficient number of times, a measurement of the resulting state
vector shall produce a solution of the problem with high probability.

The search problem is trivialy parallelizeable. Simply divide the list of $N$ 
elements into  sublists of size $N_q$ and search in the sublists. Therefore, this
problem can be adapted to the semi-quantum computer simply attributing to each
quantum node a search in a sublist.

Now let's check out what we gain with this procedure. Assuming $n_q$ qubits in the
quantum node, the list can be divided into $\frac{2^n}{2^{n_{q}}}$ sublists, 
where $n=\lg_{2}{N}$.  The quantum node searchs the sublist with cost $O(2^{n_q/2})$ 
and it is needed $\frac{2^n}{2^{n_{q}}}$ quantum nodes, or $\frac{2^n}{2^{n_{q}}}$ 
accesses to the same quantum node, to search the entire list. Therefore the cost
of this semi-quantum approach is 
\[ O(\frac{2^n}{2^{n_{q}}} \times 2^{n_{q}/2}) = O(\frac{2^n}{2^{n_{q}/2}})\,\,.\]

Note that if $n=n_q$, {\em i.e.}, if there is enough qubits to search the entire list
at once, the cost is that of the Grover's quantum search, namely, $O(2^{n/2})$. 
On the other hand, if there are no qubits at all, the cost reduces to the classical one, $O(2^n)$.
 Finally, with $n_q$ qubits available, the cost $ O(\frac{2^n}{2^{n_{q}/2}})$  is somewhere in between the classical and quantum searchs.

\section{The Discrete Fourier Transform}

Shor\cite{Shor} discovered that doing a Fourier Transform quantum mechanically
is exponentially faster than classically. This quantum Fourier Transform ($QFT$) is
the kernel of his algorithm to factorize numbers, with the notorious implication
of possibly breaking, with polynomial cost,  the most used public key cryptographic protocol, namely,
the RSA\cite{RSA}, that is based on the product of any two huge prime numbers.
It has to be noted that, in the Shor's factorization algorithm, rather than 
Fourier transforming an arbitrary quantum state, the $QFT$ is applied to a particular 
state as part of the process of factorization. Therefore, this subtle application
of the $QFT$ does not face two potentialy difficult problems, namely, preparing
an arbitrary quantum state\cite{preparation} and measuring the complex phases\cite{phase} of its Fourier Transform, which comes to be the problem we will treat
here.

Classically, the best known algorithm to calculate the Discrete Fourier Transform
of a real function at $N$ points ($DFT_N$) is the well known Fast Fourier
Transform ($FFT$), which costs $O(N\lg_2{N})$ operations. Our point here is to use
the $QFT$, which costs $O(\lg_2{N}\lg_2{N})$ quantum operations, to improve the
$FFT$ algorithm. This preamble done, now the semi-quantum approach to discrete
Fourier transform a function is described.

A real function tabulated at $N$ points can be substituted by its interpolating
polynomial 
\begin{equation}
\label{Eq5}
 X(t)=\sum_{j=0}^{N-1}x_jt^j\,\, ,
\end{equation}
which can be represented by the real vector
\begin{equation}
\label{Eq6}
 \vec{X}=(x_0, x_1, \cdots ,x_{N-1})\,\, ,
\end{equation}
whose Fourier transform is the complex vector
\begin{equation}
\label{Eq7}
\vec{y}=(y_0, y_1, \cdots ,y_{N-1})\,\, ,
\end{equation}
where
\begin{equation}
\label{Eq8}
 y_k\equiv X(\omega_N^k)=\sum_{j=0}^{N-1}x_j\omega_N^{kj}=\sum_{j=0}^{N-1}x_j\exp{(\frac{2\pi i kj}{N})}\,\, .
\end{equation}
The $\{ \omega_N^k\}$ are the $N$ complex roots of the unity.

The quantum version of the $DFT$ is as follows. Given the real vector $\vec{X}$, 
prepare the associated quantum state
\begin{equation}
\label{Eq9}
 |X\rangle = C_x\sum_{j=0}^{N-1}x_j|j\rangle\,\, ,
\end{equation}
where $C_x$ is a normalization factor and $\{|j\rangle\, , j=0,1,\cdots , N-1\}$ is an orthonormal basis.

The $DFT$ of $\vec{X}$ is the vector $\vec{Y}$, whose associated quantum state, 
written in the same basis as $ |X\rangle$, is 
\begin{equation}
\label{Eq10}
 |Y\rangle=C_y \sum_{k=0}^{N-1}y_k|k\rangle\,\, , 
\end{equation}
where $C_y$ is a normalization factor.

$|X\rangle$ and $|Y\rangle$ are related by the unitary operator $U_{QFT}$, 
\begin{equation}
\label{Eq11}
U_{QFT}|X\rangle = |Y\rangle\,\, .
\end{equation}
The action of $U_{QFT}$  on the basis states is
\begin{equation}
\label{Eq12}
 U_{QFT}|j\rangle = \frac{1}{\sqrt{N}}\sum_{k=0}^{N-1}\exp{(\frac{2\pi i kj}{N})}|k\rangle\,\, .
\end{equation}

\begin{figure}
\includegraphics{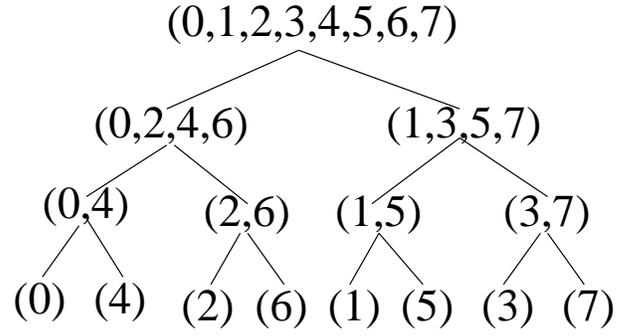}
\caption{\label{fig3} Tree representing the FFT algorithm.}
\end{figure}

The $FFT$ algorithm is based on the observation that squaring the $N$ complex roots
of the unity $\{\omega_N^k\}$ produces just $\frac{N}{2}$ distinct complex numbers
$\{\omega_{N/2}^{k}\}$. Therefore, the $DFT$ problem reduces to a recursion of identical
subproblems, with the $y_k$ evaluated as [Fig.\ref{fig3}]
\begin{equation}
\label{Eq13}
 y_k=X(w_N^k)=X_e[(w_N^k)^2]+w_N^kX_o[(w_N^k)^2]\,\, ,
\end{equation}
where
\begin{equation}
\label{Eq14}
 X_e(t)=x_0+x_2t+x_4t^2+\cdots \,\, ,
\end{equation}
\begin{equation}
\label{Eq15}
 X_o(t)=x_1+x_3t+x_5t^2+\cdots \,\, .
\end{equation}

Now assuming a quantum computer with $(n_q+1)$ qubits, let's see how to calculate
the $DFT_N$ $(N=2^n)$. For the sake of clarity, the case $n_q=2$ is explicitly 
considered, but the generalization for arbitrary $n_q$ is straightforward.

The normalized input state is 
\begin{equation}
\label{Eq16}
 |X_4\rangle = x_0|00\rangle + x_1 |01\rangle + x_2 |10\rangle + x_3 |11\rangle\,\, ,
\end{equation}
and the normalized output state is
\[ |Y_4\rangle=U_{QFT}|X_4\rangle=\]
\begin{equation}
\label{Eq17}
y_0|00\rangle + (y_{1a}+iy_{1b})|01\rangle
+ y_2|10\rangle + (y_{1a}-iy_{1b})|11\rangle\,\, ,
\end{equation}
where the ${x_k}$ and ${y_k}$ are real numbers.

\begin{figure}
\includegraphics{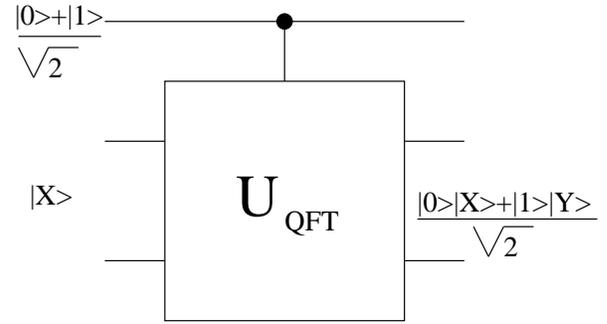}
\caption{\label{fig4} Quantum circuit to determine the Fourier phases.}
\end{figure}

In Fig.\ref{fig4}, we propose a quantum circuit for this problem. The circuit is a slight 
modification of the original $QFT$ one and includes an ancillary qubit, 
which is necessary for the proper determination of the Fourier phases.

In Tab.\ref{tab1}, we show how the Fourier phases can be measured. It involves four projections
followed by eight measurements. Then some classical operations, summarized in Tab.\ref{tab2},
are necessary to {\em rebuild} the phases. Note that the role of the ancillary qubit is to determine the sign ($+$ or $-$) of the real figures.

\begin{table}
\caption{\label{tab1} Measurement procedure to determine the Fourier phases.}
\begin{ruledtabular}
\begin{tabular}{|c|c|c|l|}
\multicolumn{2}{|c|}{STEP 1} & \multicolumn{2}{c|}{STEP 2} \\ \hline

Projection   & Result & Measurement  & Result \\
 in the basis  & &  in the basis &  \\
\hline

$|00\rangle$ & $\frac{x_0|0\rangle + y_0|1\rangle}{\sqrt{2}}$
& & \\
 & & $|1\rangle$ & $m_1\equiv\frac{|y_0|^2}{2}$ \\
 & & $\frac{|0\rangle + |1\rangle}{\sqrt{2}}$ & $m_2\equiv\frac{|x_0+y_0|^2}{2}$ \\
\hline

$|10\rangle$ &  $\frac{x_2|0\rangle + y_2|1\rangle}{\sqrt{2}}$
& & \\
 & & $|1\rangle$ & $m_3\equiv\frac{|y_2|^2}{2}$ \\
 & & $\frac{|0\rangle + |1\rangle}{\sqrt{2}}$ & $m_4\equiv\frac{|x_2+y_2|^2}{2}$ \\
\hline

$\frac{|01\rangle + |11\rangle}{\sqrt{2}}$ & $\frac{x_1+x_3}{2}|0\rangle$ & & \\
                                           & $\hspace{1cm}+y_{1a}|1\rangle$ &                            & \\
& & $|1\rangle$ & $m_5\equiv|y_{1a}|^2$ \\
& & $\frac{|0\rangle + |1\rangle}{\sqrt{2}}$ & $m_6\equiv$ \\
& &                                          & $\frac{1}{2}|\frac{x_1+x_3}{2}+y_{1a}|^2$ \\ \hline

$\frac{|01\rangle - |11\rangle}{\sqrt{2}}$ & $\frac{x_1-x_3}{2}|0\rangle$ & & \\
                                           & $\hspace{1cm} +iy_{1b}|1\rangle$              & & \\
& & $|1\rangle$ & $m_7\equiv|y_{1b}|^2$ \\
& & $\frac{|0\rangle +i |1\rangle}{\sqrt{2}}$ & $m_8\equiv$ \\
& &                                           & $\frac{1}{2}|\frac{x_1-x_3}{2}+y_{1b}|^2$ \\ 

\end{tabular}
\end{ruledtabular}
\end{table}

\begin{table}
\caption{\label{tab2} Classical operations to rebuild the Fourier phases.}

\begin{ruledtabular}
\begin{tabular}{|c|l|}

$y_0$ & If $\frac{(x_0+\sqrt{2m_1})^2}{2}\neq m_2$ \\
      & Then $y_0=-\sqrt{2m_1}$ \\
      & Else $y_0=+\sqrt{2m_1}$ \\
      & End If \\
\hline

$y_2$ & If $\frac{(x_2+\sqrt{2m_3})^2}{2}\neq m_4$ \\
      & Then $y_2=-\sqrt{2m_3}$ \\
      & Else $y_2=+\sqrt{2m_3}$ \\
      & End If \\
\hline

$y_{1a}$ & If $\frac{(\frac{x_1+x_3}{2}+\sqrt{m_5})^2}{2}\neq m_6$ \\
         & Then $y_{1a}=-\sqrt{m_5}$ \\
         & Else $y_{1a}=+\sqrt{m_5}$ \\
         & End If \\
\hline
$y_{1b}$ & If $\frac{(\frac{x_1-x_3}{2}+\sqrt{m_7})^2}{2}\neq m_8$ \\
         & Then $y_{1b}=-\sqrt{m_7}$ \\
         & Else $y_{1b}=+\sqrt{m_7}$ \\
         & End If \\
\hline

$y_1$ & $y_1=y_{1a}+iy_{1b}$ \\
\hline
$y_3$ & $y_3=y_{1a}-iy_{1b}$ \\

\end{tabular}
\end{ruledtabular}

\end{table}

Now let's see what we have gained with this semi-quantum approach to perform the
$DFT_N$. The classical part of the algorithm can still be represented by a tree, but
now with just $(n-n_q)$ levels and the operations to be done are those summarized in Tab.\ref{tab2} and Eq.\ref{Eq13}. The polynomial evaluations of Eq.\ref{Eq14} and Eq.\ref{Eq15} are avoided in the semi-quantum algorithm. Each level of the tree  [Fig.\ref{fig3}] involves $O(2^n)$ classical
operations and, therefore, the total cost is $O[(n-n_q)2^n]$. The quantum part of the
algorithm involves $\frac{2^n}{2^{n_q}}$ quantum nodes, or accesses to the same quantum
node, and each quantum node performs a $QFT$, which involves $O(n_q^2)$ quantum 
operations, resulting in a total cost of $O(n_q^22^{n-n_q})$. The access to the 
quantum node also includes an arbitrary state preparation that, according to 
Long {\em et. al.} \cite{preparation}, costs $O(n_q^22^{n_q})$. Therefore, the
total cost of the quantum part of the algorithm is (preparation +  $QFT$:) $O(n_q^22^n) + O(n_q^22^{n-n_q})$.
Summarizing, performing a Discrete Fourier Transform on the semi-quantum computer costs
(state preparation + $QFT$ + classical operations:) 
$O(n_q^22^n)+O(n_q^22^{n-n_q})+O[(n-n_q)2^n]$. 
For a fixed number of qubits ($n_q$ constant), the total cost can be rewritten as
(quantum operations + classical operations:) $O(2^n)+O(n2^n-2^n)$. This expression
shows that $O(2^n)$ classical operations are being substituted by $O(2^n)$ quantum
operations. Therefore, this semi-quantum approach is profitable only in the case that
the quantum operations are faster than the classical ones. It is reasonable  to speculate that the 
quantum operations would be {\em faster} than the classical ones,
if we remember that to determine the $2^n$ phases, we handle just $n$ qubits and
obtain real figures with  the ({\em hopefully  high}) precision of the detectors.
On the other hand, in the classical procedure, we have to handle $2^n\times n_{precision}$ bits, where $n_{precision}$
is the number of bits necessary to represent the real figures with the desired precision.
Therefore, for a given precision, the quantum procedure is always exponentially smaller
in space ({\em i.e.}, number of bits or memory) than the classical one.
It is worth  noticing that, ignoring the state preparation, the cost of the $DFT$ as a function
of the number of qubits ($n_q$) goes from the classical $O(n2^n)$, when $n_q=0$, to 
the quantum $O(n^2)$, when $n_q=n$, and for an arbitrary $n_q$, the cost is somewhere
in between the both. Therefore, if the complexity of the state preparation stage
could be improved, the semi-quantum approach would be certainly profitable.

\section{Conclusion}
In this paper, we have searched for evidences that told us if a semi-quantum computer,
which has an architecture that includes both  classical and quantum processors 
communicating classically, would have some advantage over a pure classical 
architecture. We have shown how to perform a search in a list and a Fourier Transform
in this semi-quantum computer. In the former case, we have shown that the Grover
algorithm \cite{Grover} is trivially adaptable to the semi-quantum computer and has
a performance ($O(\frac{2^n}{2^{n_q/2}})$) that is always superior to the classical
one ($O(2^n)$) and inferior to the pure quantum one ($O(2^{n/2})$). In the latter case,
we have used the Quantum Fourier Transform (QFT) algorithm \cite{Shor} ($O(n^2)$) to
improve the classical Fast Fourier Transform algorithm ($O(n2^n)$). We have shown that
using a single ancillary qubit to control the QFT transformation allows us to measure the 
Fourier phases. Due to the costly state preparation stage, we have concluded that what we
profit in the semi-quantum approach is to save $O(2^n)$ classical operations, that
involves $O(2^n)$ bits, in favor of $O(2^n)$ quantum operations, that involves just n qubits.
The expression we have obtained for the cost of this semi-quantum approach as a function of
the number of qubits in the quantum node, namely, (state preparation + QFT + classical operations:) $O(n_q^22^n)+O(n_q^22^{n-n_q})+O[(n-n_q)2^n]$, shows that, ignoring the state preparation
stage, the cost, as the number of qubits grows, ranges from the classical $O(n2^n)$ to the
quantum $O(n^2)$. On the other hand, the least cost one can hope for the arbitrary state 
preparation stage is $O(2^n)$ and, therefore, it will  allways dominate the complexity of the
semi-quantum Fourier Transform. Finally, our results suggest that a semi-quantum computer
could be an improvement over a pure classical architecture, even in the case of a small
number of qubits. Implementations of the two algorithms discussed in this paper on 
an semi-quantum computer based on optics, be quantum or classical, seems to be
relatively easy and we intend to perform such experiments to test these ideas.

\begin{acknowledgments}

Finantial support by the brazilian agencies  FAPEMIG, CAPES and CNPq. We would like to acknowledge
Maur\'{\i}cio V.B Pinheiro, Gilberto R. Medeiros, Paulo H.S. Ribeiro and Rog\'erio M. Paniago for fruitfull discussions.

\end{acknowledgments}

\end{document}